\documentclass[superscriptaddress,reprint]{revtex4-1}
\usepackage{latexsym,graphicx,amsmath,amsfonts,amssymb,verbatim}

\def\N{\mathbb{N}}

\def\d{\mathord{\mathrm{d}}}
\let\theta=\vartheta
\let\kappa=\varkappa

\def\ket#1{| #1 \rangle}
\def\bra#1{\langle #1 |}
\def\proj#1{\ket{#1}\bra{#1}}

\def\op#1#2{\ket{#1}\bra{#2}}

\def\hh{\widehat{\mathbb{H}}}
\def\one{\hat{\mathbb{I}}}

\begin{document}

\author{V\'aclav Poto\v cek}
\affiliation{SUPA School of Physics and Astronomy, University of Glasgow, Glasgow G12 8QQ, United Kingdom.}
\affiliation{Czech Technical University in Prague, Faculty of Nuclear Sciences and Physical Engineering, Department of Physics, B\v rehov\'a 7, 115 19 Praha 1, Czech Republic.}

\author{Filippo M. Miatto}
\email{miatto@gmail.com}
\affiliation{Institute for Quantum Computing and Department of Physics and Astronomy, University of Waterloo, 200 University Avenue West, Waterloo, ON N2L 3G1, Canada.}
\affiliation{Department of Physics, University of Ottawa, 150 Louis Pasteur, Ottawa, ON K1N 6N5, Canada.}

\author{Mohammad Mirhosseini}
\email{mirhosse@optics.rochester.edu}
\affiliation{The Institute of Optics, University of Rochester, 500 Joseph C. Wilson Blvd., Rochester, NY 14627, USA.}

\author{Omar S. Maga\~na-Loaiza}
\affiliation{The Institute of Optics, University of Rochester, 500 Joseph C. Wilson Blvd., Rochester, NY 14627, USA.}

\author{Andreas C. Liapis}
\affiliation{The Institute of Optics, University of Rochester, 500 Joseph C. Wilson Blvd., Rochester, NY 14627, USA.}

\author{Daniel K.\,L. Oi}
\affiliation{SUPA Department of Physics, University of Strathclyde, Glasgow G4 0NG, United Kingdom.}

\author{Robert W. Boyd}
\affiliation{Department of Physics, University of Ottawa, 150 Louis Pasteur, Ottawa, ON K1N 6N5, Canada.}
\affiliation{The Institute of Optics, University of Rochester, 500 Joseph C. Wilson Blvd., Rochester, NY 14627, USA.}
\affiliation{SUPA School of Physics and Astronomy, University of Glasgow, Glasgow G12 8QQ, United Kingdom.}

\author{John Jeffers}
\affiliation{SUPA Department of Physics, University of Strathclyde, Glasgow G4 0NG, United Kingdom.}

\title{The Quantum Hilbert Hotel}

\begin{abstract}
In 1924 David Hilbert conceived a paradoxical tale involving a hotel with an infinite number of rooms to illustrate some aspects of the mathematical notion of ``infinity''. In continuous-variable quantum mechanics we routinely make use of infinite state spaces: here we show that such a theoretical apparatus can accommodate an analog of Hilbert's hotel paradox.
We devise a protocol that, mimicking what happens to the guests of the hotel, maps the amplitudes of an infinite eigenbasis to twice their original quantum number in a coherent and deterministic manner, producing infinitely many unoccupied levels in the process. We demonstrate the feasibility of the protocol by experimentally realising it on the orbital angular momentum of a paraxial field. This new non-Gaussian operation may be exploited for example for enhancing the sensitivity of N00N states, for increasing the capacity of a channel or for multiplexing multiple channels into a single one.
\end{abstract}

\maketitle

The ``Hilbert Hotel Paradox'' demonstrates the counterintuitive nature of infinity~\cite{HilbertHotel}. The Hilbert Hotel has infinitely many rooms numbered $1,2,3,\ldots$, all of which are currently occupied. Each new visitor that arrives can be accommodated if every current guest in the hotel is asked to move up one room ($n\mapsto n+1$). Even if a countably infinite number of new guests arrives at once, they can still be accommodated if each of the existing occupants moves to twice their current room number ($n\mapsto 2n$) leaving the odd-numbered rooms free.

We may ask whether such phenomena can exist physically. One possibility is in continuous-variables systems where in principle we have infinite ladders of energy eigenstates. Previously~\cite{Oi2013}, the first of the Hilbert Hotel Paradoxes (with a single new guest) was proposed in cavity QED using the Sudarshan-Glogower bare raising operator $\hat E^{+}=\sum_{n=0}^{\infty} \op{n+1}{n}$ that shifts all the amplitudes up one level leaving the vacuum state unoccupied. Here, we show how we can implement the extended case where every second level of an infinite set of states is vacated. This can be performed coherently and deterministically, preserving all the initial state amplitudes by remapping them to twice their original levels using a short and simple sequence of instantaneous, dynamic, and adiabatic processes.

We first show how to map the eigenstate amplitudes of a infinite square potential well to twice their original level, and then we report results of a physical implementation of an analogous protocol on the Orbital Angular Momentum (OAM) eigenstates of light, where we coherently multiply any linear superposition by a fixed integer (in our case, by three). In the supplementary material we describe further details of the experiment and we show that the square well protocol can be generalised to implement a multiplication of the eigenstate numbers by any positive integer, not only by two.

\begin{figure}
\begin{center}
\includegraphics[width=3in]{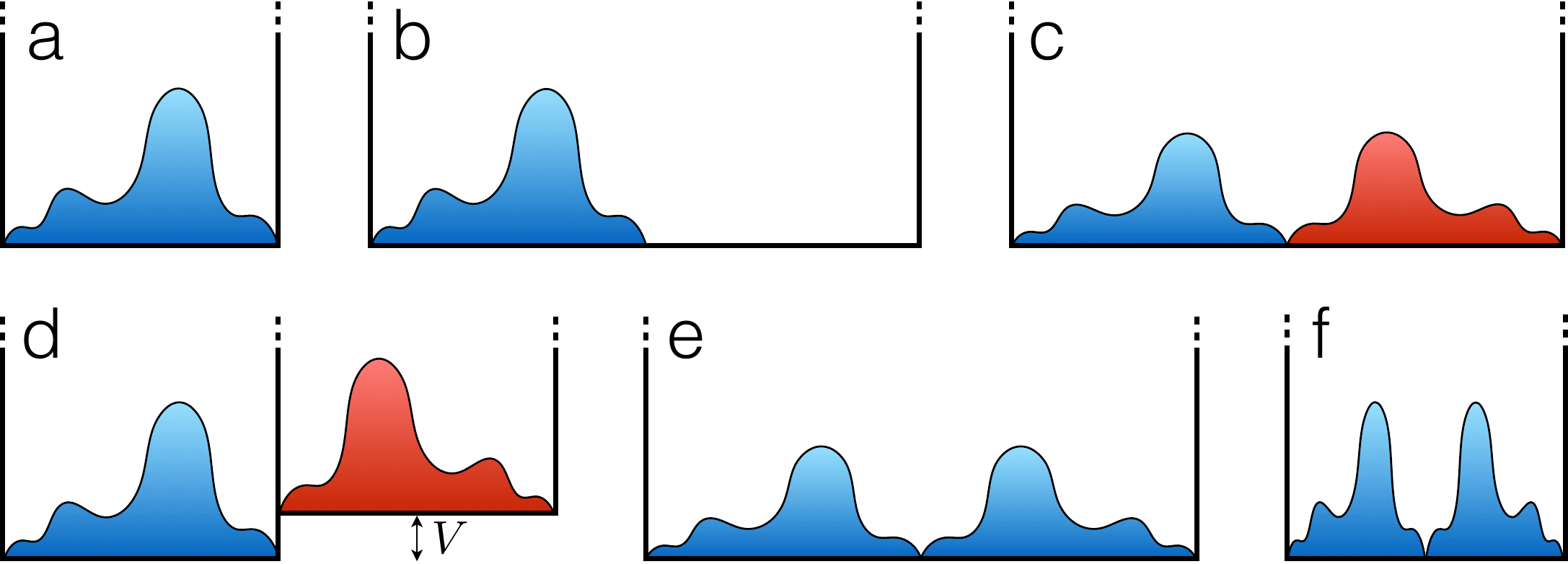}
\caption{{\bf The Hilbert Hotel Protocol.} {\bf a}, The initial state is a single  particle wavefunction $\psi(x)$ within an infinite square potential well. {\bf b}, We instantaneously expand the well to twice its original  width. The original wavefunction is not immediately changed but the eigenbasis is different. {\bf c}, We allow free evolution for a period corresponding to the original fundamental period. The wavefunction is reflected around the centre of the expanded well, with an undesired phase shift. {\bf d}, We insert an infinite  barrier in the centre (where the wavefunction is zero) to split it into two independent wells that evolve separately, an energy shift on one well corrects the relative phase. {\bf e}, After the phase correction we align the potentials and merge the two halves back together. {\bf f}, An adiabatic compression of the  well maps the eigenstates of the expanded well to those of the original well. The original wavefunction has now been halved and reflected, corresponding to the Hilbert Hotel operation $\hh_2$ being applied to the eigenstates $\psi(x)$.\label{fig:protocol}}
\end{center}
\end{figure}

Consider a quantum system with an infinite ladder of energy eigenstates bounded from below, $\{\ket{n}\}_{n=1}^\infty$. An arbitrary state can be then represented as $\ket{\psi}=\sum_{n=1}^\infty \alpha_n\ket{n}$. Our earlier work~\cite{Oi2013} has introduced the Hilbert Hotel operator $\hh_1$, transforming $\ket{\psi}$ to
\begin{equation}
\hh_1\ket{\psi}=\sum_{n=1}^\infty\alpha_n \ket{n+1}.
\end{equation}
Our new aim is to extend the toolbox by an operator $\hh_2$,
\begin{equation}
\hh_2\ket{\psi}=\sum_{n=1}^\infty\alpha_n \ket{2n},
\end{equation}
representing the second Hilbert Hotel Paradox by leaving every second energy level vacant.
Both operators are non-unitary isometries, as $\hh_j \hh_j^\dagger \ne \one$. We show that we can deterministically implement $\hh_2$ on a infinite square potential well with initial width $L$ with the following operations (Fig.~\ref{fig:protocol}):
\emph{i})~we instantaneously expand the well from $L$ to $2L$,
\emph{ii})~we let it evolve for the original fundamental period,
\emph{iii})~we divide the well into two sub-wells of width $L$ with a barrier,
\emph{iv})~we let each half-well evolve with a relative potential offset, to correct the relative phase,
\emph{v})~we merge the half-wells together into one well of width $2L$,
\emph{vi})~we adiabatically shrink the well back to width $L$.
In general, the amplitudes of an initial state can be mapped to any integer multiple ($\alpha_n\ket{n}\mapsto\alpha_{n}\ket{pn}$) using a slightly modified procedure (see supplementary material for details).

Ideally steps \emph{i}, \emph{iii} and \emph{v} should be instantaneous while step \emph{vi} should be adiabatic. The fidelity of a physical implementation will depend on the accuracy of the timing and the quality of the approximations, especially the maximum effective excitation number $n$ of the initial state in comparison to the validity regime of the Schr\"odinger equation approximation in any realistic system under consideration.

The Hilbert space of a particle in a well of width $L$ consists of the set of square-integrable functions, $L^2(0,L)$ and the free particle Hamiltonian is
\begin{align}
\hat H = -\frac{\hbar^2}{2m}  \frac{\d^2}{\d x^2},
\label{ham}
\end{align}
with boundary conditions $\psi(0) = \psi(L) = 0$. This describes a one-dimensional particle in an infinite square potential well, but it can also describe other situations, e.g. an ideal two-dimensional optical waveguide within the paraxial wave approximation. The Hamiltonian \eqref{ham} yields an infinite ladder of nondegenerate energy eigenfunctions of the form
\begin{align}
h_n(x) = \sqrt\frac2L \sin\frac{\pi n x}L, \quad n \in \N\ \mathrm{and}\ x\in(0,L),
\label{psi-n}
\end{align}
with eigenvalues $E_n=\hbar\omega_0 n^2$
where $\omega_0=\frac{\hbar\pi^2}{2mL^2}$. The desired operation $\hh_2$ transforms an initial state
$
\psi_\mathrm{in}(x) = \sum_{n=1}^\infty \alpha_n h_n(x)
$
into
\begin{align}
\psi_\mathrm{out}(x) = \sum_{n=1}^\infty \alpha_n h_{2n}(x),
\label{psi-out}
\end{align}
interleaving the amplitudes of the initial state in the energy eigenbasis with zeros.

The first step of the Hilbert Hotel protocol is to double the width of the well so the original wave function $\psi_\mathrm{in}(x)$ extends from $(0,L)$  to $(0,2L)$, filling the new interval by constant zero. We denote this extended wave function by $\psi_\mathrm{in}'(x)$ and the free Hamiltonian with the new boundary conditions $\psi(0) = \psi(2L) = 0$ by $\hat H'$. This Hamiltonian has a new set of eigenfunctions $g_n(x)$ which we use to express $\psi_\mathrm{in}'(x)=\sum_{n=1}^\infty \beta_n g_n(x)$. 
We allow $\psi_\mathrm{in}'(x)$ to evolve over a time $\tau=\frac{2\pi}{\omega_0}=\frac{mL^2}{\hbar \pi}$ into
$$\hat U'(\tau) \psi_\mathrm{in}'(x) = e^{-\frac{i \tau}\hbar  \hat H'} \psi_\mathrm{in}'(x) = \sum_{n=1}^\infty e^{-i 
\frac\pi2 n^2} \beta_n g_n(x).$$
where $e^{-i\frac\pi 2 n^2}$ is $1$ for even $n$ and $-i$ for odd $n$, thus
\begin{align}
\hat U'(\tau) \psi_\mathrm{in}'(x) &= \sum_{m=1}^\infty \beta_{2m} g_{2m}(x) - i\sum_{m=1}^\infty 
\beta_{2m-1} g_{2m-1}(x)\nonumber\\
 &= \left( \frac{1-i}2 \hat I - \frac{1+i}2 \hat R \right) \psi_\mathrm{in}'(x),
\label{propagator-2}
\end{align}
where $\hat I$ is the identity operator and $\hat R = (-1)^{\hat m+1}$ the mirror reflection (or parity) operator. Therefore, after step \emph{ii} we have (up to a global phase factor) the state
\begin{align}
\hat U(\tau)\psi_\mathrm{in}'(x) = \frac1{\sqrt2} 
\begin{cases}
\psi_\mathrm{in}(x) & x \in (0,L) \\
-i\psi_\mathrm{in}(2L-x) & x \in (L,2L).
\end{cases}
\label{replication}
\end{align}

This resembles the point symmetry extension of $\psi_\mathrm{in}(x)$ to $(0,2L)$ but the phase factor in $(L,2L)$ needs to be corrected. Steps \emph{iii}, \emph{iv} and \emph{v} remove the undesired $i$ factor while preventing cross-talk between the two sub-wells. After splitting the interval $(0,2L)$ in two, each part will evolve separately under the Hamiltonian
\begin{align}
\hat H_\mathrm{offset}= -\frac{\hbar^2}{2m} \frac{\d^2}{\d x^2} + V,
\end{align}
with appropriate boundary conditions. The two halves can be phase-matched by applying potentials $V=0$ in $(0,L)$ and $V = \hbar\omega_0/4$ in $(L,2L)$ for a time $\tau = 2\pi/\omega_0$.

After removing the barrier (step \emph{v}), the wave function of the system becomes
$$\psi_\mathrm{phase}(x) = \frac1{\sqrt2} \begin{cases}
\psi_\mathrm{in}(x) & x \in (0,L) \\
-\psi_\mathrm{in}(2L-x) & x \in (L,2L).
\end{cases}$$
Substituting for $h_n(x)$ from \eqref{psi-n}, we find that both branches allow for a common analytic expression,
 as the domain of $g_n(x)$ is twice that of $h_n(x)$:
$$\psi_\mathrm{phase}(x) = \sqrt\frac1L \sum_{n=1}^\infty \alpha_n \sin\frac{\pi 
n x}L = \sum_{n=1}^\infty \alpha_n g_{2n}(x).$$

The final step is an adiabatic compression of the well back to its original width $L$. Up to a relative phase due to free evolution, which can be corrected by matching the total time of the evolution to an integer number of full revolutions of the running eigenbasis, this adiabatically transforms the basis states $g_n(x)$ into $h_n(x)$ of the same $n$, keeping coherent superpositions intact. This shows the resulting state is indeed \eqref{psi-out}.

\begin{widetext}
\begin{figure*}
\begin{center}
\includegraphics[width=6in]{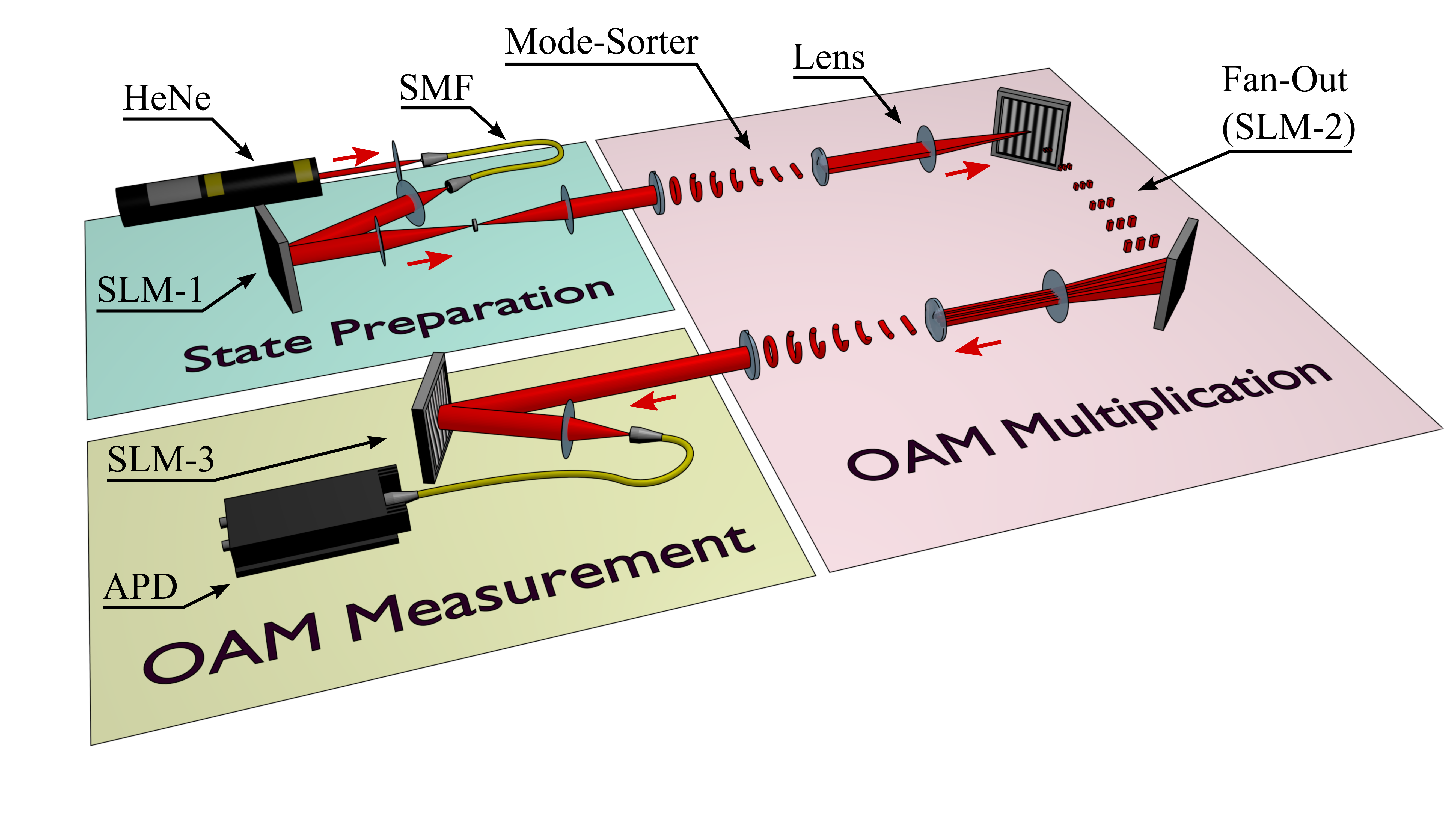}
\caption{\label{HHsetup}{\bf Experimental Schematic.} A spatially filtered and collimated HeNe laser beam is directed onto a phase-only spatial light modulator (SLM-1) to generate the desired combination of input OAM eigenmodes. The beam is then sent through a pair of machined polymer refractive elements that comprise the first OAM sorter. The optical field at the output plane of this sorter is imaged onto the top half of a second SLM implementing a fan-out grating. The fan-out was set to produce three copies of the beam, resulting in a $\times 3$ multiplication of the OAM quantum number of each mode. The Fourier plane of this grating is imaged onto the bottom half of the same SLM displaying the appropriate hologram to correct for the relative phase between the three copies. The three copies are then demagnified by a cylindrical lens and injected through a second OAM sorter operated in reverse. We measured the OAM spectrum of light at the output of the multiplier using a series of projective measurements for various values of $\ell$, which were implemented using a third SLM and a single-mode-fibre-coupled avalanche photodiode (APD).}
\end{center}
\end{figure*}
\end{widetext}

The crucial step in the Hilbert Hotel operation is the coherent mapping $|n\rangle\mapsto |pn\rangle$ (for $p\in \mathbb{Z}^+$) on a countably infinite set of basis states $\{|n\rangle\}$, as described above. Instead of a particle in an infinite square potential well, we can use systems that share important characteristics in order to perform analogous operations. In our experimental realisation (Fig.~\ref{HHsetup}) we choose the set of OAM eigenstates of a beam of light, denoted by $\ket{\ell}$, and the coherent multiplication makes use of two well-known optical devices in a novel configuration: an OAM sorter and a ``fan-out'' refractive coherent beam copier~\cite{Gale1992, Romero2007}.

The OAM multiplier has four steps: \emph{i}) unwrapping the initial azimuthal phase ring into a linear phase ramp with a polar-to-cartesian mapping, \emph{ii}) branching out new copies of the linearised field and correcting their relative phase with a suitable grating, \emph{iii}) demagnifying the juxtaposed copies with a cylindrical lens, and \emph{iv}) wrapping the resulting field back to polar coordinates. The combination of these four steps amounts to the transformation:
\begin{align}
\sum_\ell c_\ell|\ell\rangle\ \mapsto\ \sum_\ell c_\ell |p\ell\rangle,
\end{align}
where $p$ is the number of copies produced in step \emph{ii}.
The first step is achieved by way of an OAM sorter~\cite{Berkhout_PRL2010,Mirhosseini_NComms2013}, which unwraps any OAM mode into a linear gradient (and therefore it turns a combination of OAM modes into a combination of linear gradients) by way of an extremely astigmatic lens $\phi_1$ followed by a phase-correcting element $\phi_2$, which effectively stops the unwrapping after the transformation is complete.
These two elements can be described by the phase delay that they impose on the incoming field as a function of position:
\begin{align}
\phi_1(x,y)&=a\frac{2\pi}{\lambda f}\left(y \arctan\frac{y}{x}-x\log\frac{\sqrt{x^2+y^2}}{b}+x\right),\\
\phi_2(u,v)&=-ab\frac{2\pi}{\lambda f}\exp\left(\frac{-u}{a}\right)\cos\left(\frac{v}{a}\right),
\end{align}
where $f$ is the focal length of the Fourier lens connecting near-field and far-field, $\lambda$ is the wavelength of the light beam, and the free parameters $a$ and $b$ determine the scaling and position of the transformation in the Fourier plane of coordinates $u$ and $v$.

At this point we produce equal-weighted copies of the unwrapped phase ramp using a fan-out element by way of a suitable 1D phase grating on the far field. It is crucial that the copies have the same intensity in order to obtain the desired OAM modes at the end of the process. In our experiment, the fan-out grating produces three copies and the equation describing the phase delay of the grating as a function of position in the far field is
\begin{align}
\varphi(x,y)=\arctan[2\mu\cos(x)],
\end{align}
where $\mu\approx 1.32859$. Such a phase mask does not depend on the $y$ coordinate, as we are copying a linear field. This grating is displayed on a spatial light modulator (SLM), so the output of the sorter needs to be Fourier transformed onto the fan-out SLM with a $2f$ system, followed by another $2f$ system which images it through a second sorter operated \emph{in reverse}. In order to wrap the field back correctly without leaving wide gaps or without wrapping more than necessary, we use a cylindrical lens to demagnify the horizontal cartesian coordinate before the beam enters the reverse-sorter. Exploiting the flexibility of SLMs, we achieve this by adding the phase of a cylindrical lens directly on top of the fan-out grating.

\begin{figure}
\begin{center}
\includegraphics[width=3in]{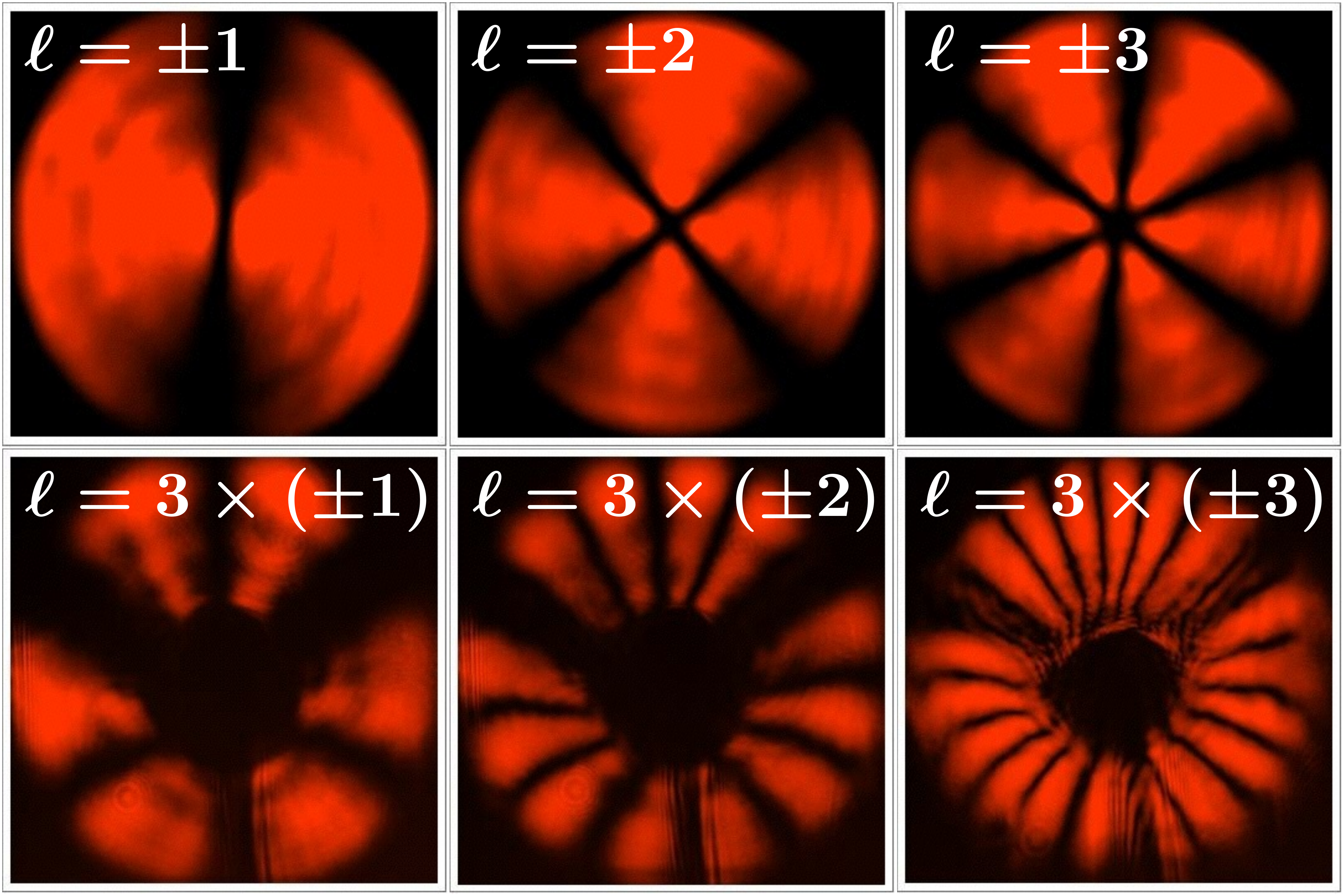}
\caption{\label{petals} {\bf Coherent OAM multiplication.} Top row: Near field of input coherent superpositions. Bottom row: Tripled output states. The number of petals is $6|\ell|$, as expected from a coherent operation.}
\end{center}
\end{figure}

In the first part of our experiment we test the coherence of the protocol, i.e. its ability to preserve superpositions. To do this, we generate balanced superpositions of $+\ell$ and $-\ell$, with $\ell$ ranging from 1 to 3. Such initial modes display $2|\ell|$ maxima, or ``petals''. We feed them to the multiplier (here set to multiply by $p=3$) and a successful protocol results in $6|\ell|$ petals with high visibility at the output, as can be seen in Fig.~\ref{petals}.

\begin{figure}
\begin{center}
\includegraphics[width=3in]{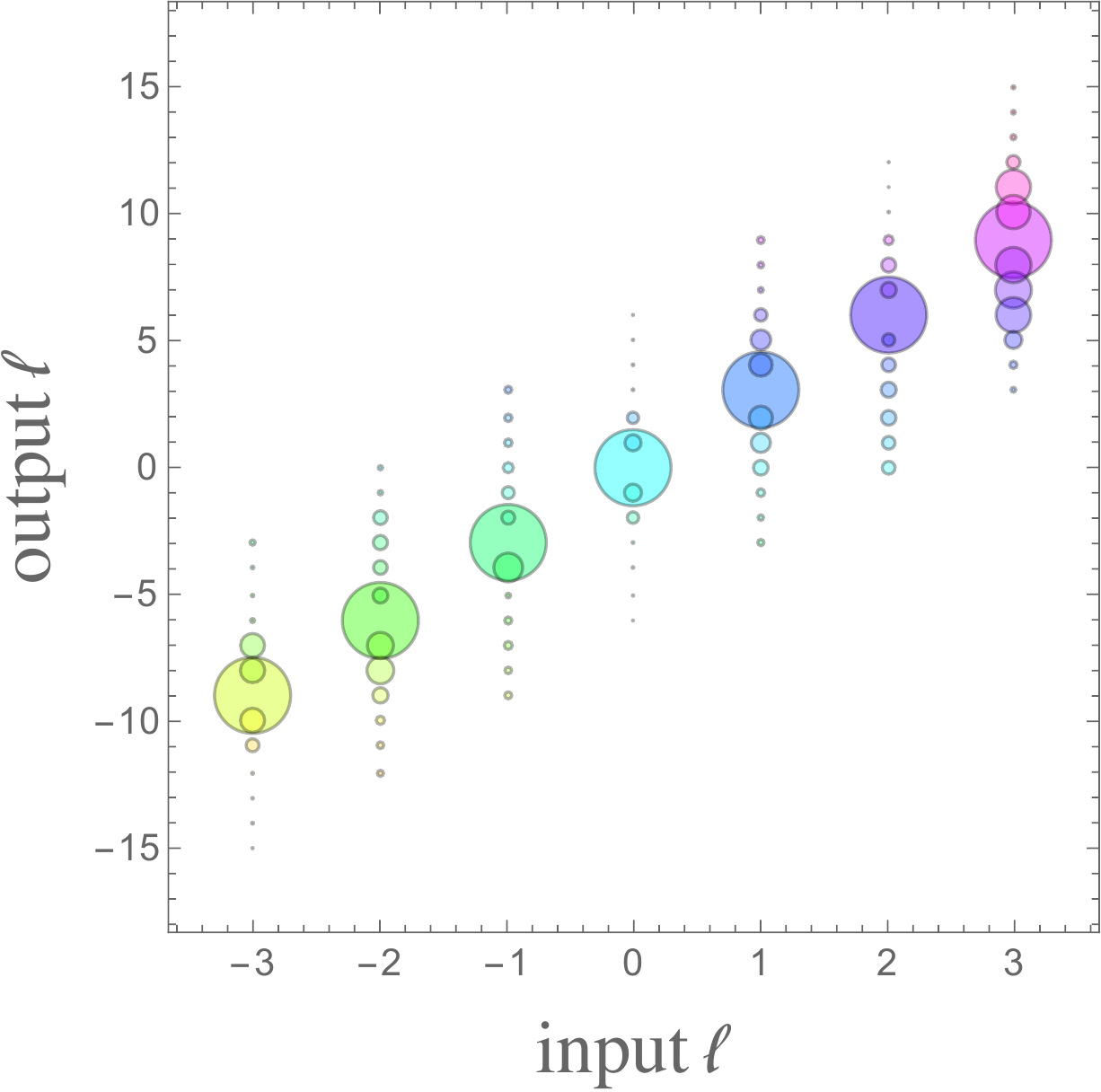}
\caption{\label{DataAnalysis}{\bf OAM multiplication performance.} For each input eigenmode we measure the composition of the multiplied output. Circle size is linearly proportional to the overlap with the output modes. As can be seen, the small leakage onto the neighbouring output modes is contained within a few adjacent modes. A sufficiently distant input superposition such as $|3\rangle+|-3\rangle$ would maintain an effective orthogonality.}
\end{center}
\end{figure}

In the second part of our experiment we assess the accuracy of the protocol 
by measuring the leakage onto neighbouring OAM eigenmodes. To do this, we multiply single OAM eigenmodes by $p=3$ and projectively measure the OAM spectrum of the output. The results show that the overlap decays quickly enough for suitably distant superpositions to maintain their orthogonality (Fig.~\ref{DataAnalysis}). For instance, the superposition $|3\rangle+|-3\rangle$ which ideally maps to $|9\rangle+|-9\rangle$, was mapped to a superposition of modes, peaked on $\ell=\pm9$, but nevertheless with negligible cross-talk (details in supplementary material).

In summary, we showed how to implement the Hilbert Hotel ``paradox'', where the rooms of the hotel are the excitation modes of an infinite square potential well. We then reported the successful implementation of the core step of the operation (the coherent multiplication of the basis states of a countably infinite basis) on the OAM eigenmodes of a paraxial beam of light. We show that the operation is coherent and that even in our proof-of-principle experiment, the multiplication of sufficiently distant modes can be performed with negligible overlap.
Mode multiplication could be implemented also in other quantum systems, such as BECs in a box potential with predicted Talbot carpet features, though nonlinear interactions may spoil the ideal free particle expansion required for perfect wavefunction mirroring~\cite{RKSR2001}.  Nonetheless, we note that this idea could be used to enhance several state production schemes without the need to modify the existing apparatuses, because it can act as an extension. For instance, it could prove useful in quantum and classical information processing as a means of multiplexing an arbitrary number of input channels into a single output channel, or to enhance the sensitivity of systems that use N00N states, or to distribute ordered gaps in the spectral profile of a state.

\section*{Acknowledgements}
This work was supported by the Canada Excellence Research Chairs (CERC) Program, the Natural Sciences and Engineering Research Council of Canada (NSERC) and the UK EPSRC. O.S.M.L. acknowledges support from CONACyT and the Mexican Secretaria de Educacion Publica (SEP). DKLO, JJ, and VP acknowledge QUISCO.

\section*{Author contributions}
V.P., F.M.M., D.K.L.O. and J.J. developed the theory. F.M.M., M.M., O.S.M.L., A.C.L. and R.W.B. conceived the experiment. M.M. O.S.M.L. and A.C.L. carried out the experiment. F.M.M. performed the data analysis. All authors contributed to writing the paper.\\

\section*{Competing financial interests}
The authors declare no competing financial interests.

\end{document}